\documentclass[preprint,showpacs,preprintnumbers,amsmath,amssymb]{revtex4}

\usepackage{graphicx}

\newcommand{\be}{\hbox{\normalsize e}}

\begin{document}

\title{On the Superradiant Phase in Field-Matter Interactions}

\author{O. Casta\~nos, E. Nahmad-Achar, R. L\'opez-Pe\~na, and J. G. Hirsch}

\affiliation{Instituto de Ciencias Nucleares, Universidad Nacional
Aut\'onoma de M\'exico, Apdo. Postal 70-543 M\'exico 04510 D.F.\\ \\}


\begin{abstract}

We show that semi-classical states adapted to the 
symmetry of the Hamiltonian are an excellent approximation to the exact quantum solution of the ground and first excited states of the Dicke model. Their overlap to the exact quantum states is very close to $1$ except in a close vicinity of the quantum phase transition. Furthermore, they have analytic forms in terms of the model parameters and allow us to
calculate analytically the expectation values of field and matter
observables. Some of these differ considerably from results obtained via the standard coherent states, and by means of Holstein-Primakoff series expansion of the Dicke Hamiltonian. Comparison with exact solutions obtained numerically support our results. In particular, it is shown that the expectation values of the number of photons and of the number of excited atoms have no singularities at the phase transition. We comment on why other authors have previously found otherwise.
	
\end{abstract}


\maketitle


\section{Introduction}

The field-matter interaction is commonly described by the Dicke
Hamiltonian~\cite{dicke}, which considers $N$ two-level atoms with
energy separation equal to $\hbar\tilde{\omega}_{A}$ immersed in an
external one-mode electromagnetic field of frequency
$\tilde{\omega}_{F}$ inside a cavity. This Hamiltonian comes from the
standard quantization of the multipolar form of many atoms interacting
with classical radiation~\cite{power} in the long wave
approximation. 
The Dicke model was established exploiting the analogy with the
spin. It is considered a multiatom generalization of the
Jaynes--Cummings model~\cite{jaynes}, a completely soluble quantum Hamiltonian of a two-level atom in a one mode electromagnetic field. The Tavis--Cummings
model~\cite{tavis} is also a many body extension of the Jaynes--Cummings
model, which considers the rotating-wave approximation.

A very important feature of the Dicke Hamiltonian discovered by Hepp
and Lieb~\cite{hepp} is the presence of a phase transition from the
normal to the superradiant behaviour. The latter, introduced by Dicke
to describe coherent radiation, is a collective effect involving all
$N$ atoms in the sample, where the decay rate is proportional to $N^2$
instead of $N$, the expected result for independent atom emission. While this
transition to the superradiant phase has been much debated in the 
literature~\cite{wodk,birula,siva,casta3}, due to the smallness of
the field-matter coupling strength, recent experimental results
indicate that it can actually be observed~\cite{dimer,baumann,nagy}.

In this contribution we use the tensorial product of $SU(2)$- and
$HW(1)$-coherent states as a trial state to calculate the
expectation value of the Hamiltonian and construct its energy
surface. This is a function that depends on four variables
$(q,p,\theta, \phi)$ which define the coherent states, and two essential
parameters $(\omega_A,\gamma)$ which are associated to the energy
separation of the two-level atoms and the coupling strength between
the atoms and the field, respectively. By studying the stability
properties of the critical points of this energy surface we obtain
the best variational wavefunction and a separatrix which divides
the parameter space in two regions: normal and superradiant
behaviour, according to $\gamma < \gamma_c $ or $\gamma > \gamma_c $,
respectively, with $\gamma_c = \sqrt{\omega_A/4}$.

The Dicke Hamiltonian is invariant under transformations of the cyclic
group $C_{2}$, a symmetry which is not preserved by the trial
state. The symmetry can be restored by acting with the
projectors of the symmetric and antisymmetric representations of the
cyclic group $C_{2}$.
We determine the statistical properties of field and matter
observables in the restored variational states
$\vert\Psi_{P}\rangle$ in the superradiant region. The comparison with
the same observables evaluated from the exact diagonalization of the
Hamiltonian at finite number of atoms $N$ exhibits an excellent
agreement between them, which improves as $N$ becomes larger~\cite{rapcom}. The
fidelity between both types of states is also evaluated; we
establish that the fidelity in the superradiant regime is very close
to unity, dropping fast in a small neighbourhood of the separatrix. The
great advantage of working with the restored-symmetry states is that 
they allow us to write down analytic expressions for all the expectation
values of field and matter observables, as well as with important
mathematical relations. These analytic expressions are exact in the
thermodynamic limit, when the number of particles goes to infinity.

The Dicke Hamiltonian has been studied thoroughly by Emary and
Brandes~\cite{emary} with a variational procedure in terms of two
$HW(1)$-coherent states. One of them is related to the radiation field
degrees of freedom and the other arises from a Holstein-Primakoff
(H-P) realization of the $SU(2)$ generators~\cite{holstein}. They obtain
the appropriate relation between the parameters of the Dicke Hamiltonian
to describe the phase transition, as well as a description for the
normal phase in the thermodynamic limit. However, they consider
in the superradiant region, an approximated Dicke Hamiltonian coming from a series expansion of the H-P realization
truncated
to second order in terms of the ratio between the number of
excited atoms over the total number of atoms, assuming that
is a very small quantity. Under this assumption Nagy et al.~\cite{nagy}
found singularities at $\gamma=\gamma_{c}$ in the expectation values of the
number of photons and the number of excited atoms, and related them with
the divergences in their corresponding fluctuations~\cite{emary}.

In this contribution we extend and provide details of the calculations presented in~\cite{rapcom}. The expectation values of the mentioned
observables are calculated by means of the symmetry adapted variational states and find
that there is in fact a divergence of the expectation values, but that this remains
throughout the superradiant region, contrary to the published results:
there is no singularity. The expectation value of the number of photons
$\langle\hat{a}^{\dagger}\hat{a}\rangle$
and the squared
fluctuation of the first quadrature of the electromagnetic field
$(\Delta\hat{q})^{2}$ are proportional to the number of atoms at the
phase transition and in the superradiant regime, while the squared
fluctuation $(\Delta\hat{J}_{x})^{2}$ is proportional to $N^{2}$ in
the same region.
Comparison with the exact numerical solution supports our results.
This could have consequences in studies of entanglement and chaos in the superradiant region.

\section{Dicke Hamiltonian and Symmetry Adapted States}

The Dicke Hamiltonian involves the collective interaction of $N$
two-level atoms with energy separation $\hbar\tilde{\omega}_{A}$ with
a one-mode radiation field of frequency $\tilde{\omega}_{F}$ in the
long wavelenght limit. It has the form
   \begin{equation}
      H_{D}=\hat{a}^{\dagger}\hat{a}+\omega_{A}\hat{J}_{z}
      +\frac{\gamma}{\sqrt{N}}\left(\hat{a}^{\dagger}+\hat{a}\right)
      \left(\hat{J}_{+}+\hat{J}_{-}\right)\ ,
      \label{eq001}
   \end{equation}
where $\omega_{A}\equiv\tilde{\omega}_{A}/\tilde{\omega}_{F}$ is given
in units of the frequency of the field, and
$\gamma=\tilde\gamma/\tilde{\omega}_F=\sqrt{( 2\pi\varrho/\hbar \,\tilde{\omega}_{F})}\,
\vec{d}_{ba}\cdot\vec{e}_{P}$, is the coupling parameter between the matter and field. The  
$\varrho$ denotes the density of atoms in the quantization volume, $\vec{d}_{ba}$ is the excitation matrix element of the electric dipole operator of a single atom,
and $\vec{e}_{P}$ the polarization vector. The operators
$\hat{a}^{\dagger},\,\hat{a}$ denote the one-mode creation and annihilation
photon operators; $\hat{J}_{z}$ the atomic relative population operator;
and $\hat{J}_{\pm}$ the atomic transition operators.
We will consider completely symmetric states for which we have
$N=2\,j$.

Later we will find it convenient to divide the Dicke Hamiltonian
by the total number of particles, having in this way an intensive
Hamiltonian operator. In order to study the thermodynamic limit
one would take $N\rightarrow\infty$.

To find the symmetries of the Hamiltonian, one considers the
unitary transformation
$\hat{U}=\exp\left(i\phi_{0}\hat{\Lambda}\right)$ with
$\hat{\Lambda}=\hat{a}^{\dagger}\hat{a}+\hat{J}_{z}
+\sqrt{\hat{\vec{J}}^{2}+1/4}-1/2$
denoting the excitation number operator. One can show that
	\begin{equation}
                \hat{U}\,\hat{a}\,\hat{U}^{\dagger}=e^{-i\phi_{0}}\,\hat{a}\ ,
                \qquad
                \hat{U}\,\hat{J}_{+}\,\hat{U}^{\dagger}
                =e^{-i\phi_{0}}\,\hat{J}_{+}\ ,
                \label{eq009}
        \end{equation}
with the corresponding hermitean conjugated relations. Substituting
these results into the expression for the Dicke Hamiltonian one has
	\begin{equation}
                \hat{U}\,H_{D}\,\hat{U}^{\dagger}=\hat{a}^{\dagger}\hat{a}
                +\omega_{A}\hat{J}_{z}+\frac{\gamma}{\sqrt{N}}
                \left(\hat{a}^{\dagger}\,\hat{J}_{-}+\hat{a}\,\hat{J}_{+}\right)
                +\frac{\gamma}{\sqrt{N}}
                \left(e^{-2i\phi_{0}}\,\hat{a}^{\dagger}\,\hat{J}_{+}+
                e^{2i\phi_{0}}\,\hat{a}\,\hat{J}_{-}\right)\ .
                \label{eq010}
        \end{equation}
If the counter-rotating term (last term of the
expression) is neglected, one recovers the Tavis-Cummings model
and it is immediate that it is invariant under any rotation with
arbitrary $\phi_{0}$. Their symmetrized solutions have been thoroughly studied
in~\cite{ocasta1,ocasta2}.
For the Dicke Hamiltonian, it is necessary to
restrict to rotations by an angle $\phi_{0}=0,\,\pi$ for the Hamiltonian
to remain invariant. Thus the invariance group for the
Dicke model is ${\cal C}_{2}=\left\{I,\,\be^{i\,\pi\,\hat{\Lambda}}\right\}$.

The projection operators for this group are
	\begin{equation}
                \hat{P}_{\pm}=\frac{1}{2}\left(I \pm
                \be^{i\,\pi\,\hat{\Lambda}}
                \right)\ ,
                \label{eq011}
   \end{equation}
which allow us to write the eigenfunctions of the Hamiltonian displaying
its symmetry explicitly. In terms of the eigenstates $\vert\nu\rangle$ of the photon number
operator, and the square of the angular momentum and its
projection in the z-axis $\vert j,\, m\rangle$, they may be expressed as
	\begin{equation}
                \vert\phi^{k}_{j;\pm}\rangle
                =\sum_{\lambda=0}^{\infty} \
                \sum_{\nu=\max\{0,\lambda-2j\}}^{\lambda}
                \tfrac{ \left(1  \pm (-1)^\lambda \right)}{2}\
                c^{k}_{\lambda,\,\nu} \
                \vert\nu\rangle\otimes\vert j,\,\lambda-j-\nu\rangle\ ,
                \label{eq012}
   \end{equation}
where $\vert\phi^{k}_{j; +}\rangle$ contains only even values of
$\lambda=\nu+j+m$, the eigenvalues of $\hat{\Lambda}$, while
$\vert\phi^{k}_{j; -}\rangle$ contains the odd
ones. The index $k$ denotes the state number of the even and odd
solutions. In practice one uses a maximum value for $\lambda$ which guaranteed the
convergence in the energy eigenvalues of the ground and first
excited states. The coefficients $c^{k}_{\lambda,\,\nu}$ can be
obtained from the diagonalization of the Hamiltonian matrix, whose
dimension is
	\begin{displaymath}
                d=\left\{\begin{array}{r@{\quad,\qquad}l}
                \frac{1}{2}\left(\lambda_{\max}+1\right)
                \left(\lambda_{\max}+2\right)&\lambda_{\max}\le 2j\ ;\\
                \left(2j+1\right)\left(\lambda_{\max}-j+1\right)&
                \lambda_{\max}\ge 2j\ ,
                \end{array}\right.
        \end{displaymath}
in the basis states without definite parity of $\lambda$. This matrix,
in the symmetry adapted basis states, breaks into two pieces of even
and odd parity respectively. For $\lambda\geq 2 j$, the dimensions of the matrices are
	\begin{displaymath}
                d_{+}=(j+1)^{2}+(2j+1)(s_{+}-j)\, ,\quad
                d_{-}=j(j+1)+(2j+1)(s_{-}-j)\, ,
        \end{displaymath}
for integer $j$, with $s_+=\lfloor\lambda_{max}/2\rfloor$ and $s_-=\lfloor(\lambda_{max}+1)/2\rfloor$. For $\lambda < 2 j$, they are $d_+=(s_+ +1)^2$ and $d_-=s_-(s_- +1)$. Similar
expressions can be obtained for half integer $j$.

\section{Energy Surface and Critical Points}

The energy surface is found by taking the 
expectation value of the Dicke Hamiltonian  with respect to
the tensorial product of coherent states of Heisenberg--Weyl and $SU(2)$ groups
$\vert\alpha\rangle\otimes\vert\zeta\rangle\ $, given
by~\cite{hecht,gilmore1972}
   \begin{eqnarray*}
      \vert\alpha\rangle&=&\exp\left(-\left|\alpha\right|^{2}/2\right)\,
      \sum_{\nu=0}^{\infty} \frac{\alpha^{\nu}}{\sqrt{\nu!}}
      \,\vert\nu\rangle\ ,\\
      \vert\zeta\rangle&=&\frac{1}{\left(1+
      \left|\zeta\right|^{2}\right)^{j}}\,\sum_{m=-j}^{j}
      \binom{2j}{j+m}^{1/2}\,\zeta^{j+m}\,\vert j,\,m\rangle\ ,
   \end{eqnarray*}
where the parameters $\alpha$ and $\zeta$ are complex numbers.
Using the Ritz variational principle one may find the best variational approximation to the ground state energy of the system and its corresponding eigenstate.

To calculate the expectation values of the field and matter
observables it is convenient to find the representation of the
angular momentum and Weyl generators with respect to the tensorial
product $\vert\alpha\rangle\otimes\vert\zeta\rangle\ $~\cite{hecht}.
For example, for the annihilation photon operator we consider the
matrix element
	\begin{eqnarray}
	\langle \alpha \vert \hat{a} \vert \psi \rangle &=& 
	\be^{-\tfrac{\vert \alpha\vert^2 }{2}} \, 
	\langle 0 \vert \be^{\alpha^* \, \hat{a}} \ \hat{a} \vert \psi \rangle 
	\nonumber \\
	&=& \left( \frac{\partial \ \ }{\partial \alpha^*} +
            \frac{\alpha}{2} \right)
	\,\langle \alpha\vert \psi \rangle\ ,
	\end{eqnarray}
where $\vert \psi \rangle$ denotes an arbitrary state. We proceed in a
similar form for the other observables. The representations
of the annihilation and creation photon operators are thus given by  
	\[
	\hat{a} \rightarrow \frac{\partial \ \ }{\partial \alpha^*} +
	\frac{\alpha}{2} \, ,  \quad \hat{a}^\dagger \rightarrow \alpha^* \, ,
	\]
and for the matter observables we have
       \begin{eqnarray}
	\hat{J}_z & \rightarrow & -\frac{j}{1 + \vert \zeta\vert^2} +
	\zeta^* \, \frac{\partial \ \ }{\partial \zeta^*} \, , \nonumber \\
	\hat{J}_+ & \rightarrow & j \, \zeta^* 
	\frac{2 + \vert \zeta\vert^2}{1 + \vert \zeta\vert^2} +
	(\zeta^*)^2 \, \frac{\partial \ \ }{\partial \zeta^*} \, , \nonumber \\
	\hat{J}_- & \rightarrow & \frac{j \, \zeta}{1 + \vert
            \zeta\vert^{2}}+
	\frac{\partial \ \ }{\partial \zeta^*} \, . \nonumber
        \end{eqnarray}
The energy surface can be then calculated straightforwardly~\cite{ocasta1,ocasta2}
   \begin{eqnarray}
      {\cal H}(\alpha,\,\zeta)&\equiv&
      \langle\alpha\vert\otimes\langle\zeta\vert
      H_{D}\vert\alpha\rangle\otimes\vert\zeta\rangle\nonumber\\
      &=&\frac{1}{2}\left(p^{2}+q^{2}\right)-j \,
      \omega_{A}\,\cos\theta+2\sqrt{j}\gamma\,q\,\sin\theta\,
      \cos\phi\ .
      \label{eq003}
   \end{eqnarray}
In this expression we take the harmonic oscillator realization
for the field part and the stereographic projection for the angular
momentum part,
	\begin{equation}
		\alpha=\frac{1}{\sqrt{2}}\left(q+i\,p\right)\ ,
                \qquad\zeta=\be^{-i\,\phi}\,\tan\frac{\theta}{2}\ ,
		\label{dzeta}
	\end{equation}
where $(q,p)$ correspond to the expectation values of the
quadratures of the field, and $(\theta,\phi)$ determine a
point on the Bloch sphere.

The minima and degenerate critical points are obtained by means of the
catastrophe formalism~\cite{gilmore3}. By calculating the Hessian of
the energy surface, we see that when $\gamma_{c}^{2}=\omega_{A}/4$ the
critical points degenerate and, for that value of the field-matter
coupling, the phase transition from the normal to the superradiant
behaviour of the atoms takes place.

The critical points which minimize ${\cal H}$ are given by
	\begin{equation}\label{criticos}
		\begin{array}{lllll}
		\theta_{c}=0\,,& q_{c}=0\, ,&p_{c}=0\, ,&\hbox{
                for } \vert\gamma\vert < \gamma_c \, ,\\
		\theta_{c}=\arccos(\gamma_c/\gamma)^{2}
		\, ,&q_{c}=-2\,\sqrt{j}\,\gamma\,
		\sqrt{1-(\gamma_c/\gamma)^{4}} 
		\cos{\phi_c}\, ,&p_{c}=0\, ,&\hbox{ for
                }\vert\gamma\vert > \gamma_c\ .
		\end{array} 
	\end{equation}
The last column of this array shows the conditions in parameter space
to guarantee that they constitute a minimum of the energy surface. The first
row describes the minimum critical points for the normal phase, while
the second row describes those for the superradiant regime. In this
last case one has $\phi_{c}=0,\,\pi$. (If one were to consider a case in
which $\omega_{A}<0$, we would then have
$\gamma_{c}^{2}=-\omega_{A}/4$
and the critical points for the normal phase would be
$\theta_{c}=\pi,\ q_{c}=0, \ p_{c}=0$, for $\vert\gamma\vert <
\gamma_c$). It is convenient to work with the variable $x=
\gamma/\gamma_c$ in terms
of which the energy values for the minima just described are
	\begin{equation}
	    E_{\hbox{normal}}=-2N\,\gamma_{c}^{2}\ ,\quad
                E_{\hbox{superradiant}}=-N\,\gamma_{c}^{2}\,x^{2}\,\left(
                1+x^{-4}\right)\ .
                \label{ecoherente}
	\end{equation}

In a similar form, the expectation values of $\hat{\Lambda}$ at the
minima are
	\begin{equation}
	    \lambda_{\hbox{normal}}=0\, ,
                \quad\lambda_{\hbox{superradiant}}=\tfrac{N}{2}\,\left[1
                -x^{-2}+2\,\gamma_{c}^{2}\,x^{2}\,\left(1-x^{-4}\right)
                \right]\ .
                \label{eq007}
	\end{equation}
The fluctuations $\Delta\hat\Lambda$ are zero in the normal region,
and in the superradiant phase take the form
	\begin{equation}
        \Delta\hat{\Lambda}=\sqrt{\tfrac{N}{2}
	\left(\tfrac{1}{2}+2\,\gamma^{2}_c \, x^2 \right)\left(1-x^{-4}
	\right)} \ .
        \end{equation}
To study the statistical properties of the variational states in the
superradiant regime we calculate the expectation values of linear
matter and field observables with respect to the tensorial product of
coherent states, as well as their fluctuations. The results are given
in the left side of Table \ref{tab1}. Notice that we have introduced
the quadratures operators $\hat{q}$ and $\hat{p}$ of the electromagnetic field and they can be written in terms of the creation and annihilation photon operators.

\section{Symmetry-Adapted Coherent States}

To build variational states which preserve the symmetry of the Dicke
Hamiltonian we apply the projectors of even and odd
parity~(\ref{eq011}) to the coherent
states~$\vert\alpha\rangle\otimes\vert\zeta\rangle$, to obtain
	\begin{equation}
		|\alpha,\,\zeta \rangle_{\pm}={\cal N}_{\pm}\Big(
		\vert\alpha\rangle\otimes\vert\zeta\rangle\pm\
		\vert-\alpha\rangle\otimes\vert-\zeta\rangle\Big)\ ,
		\label{eq013}
	\end{equation}
where the normalization factors ${\cal N}_{\pm}$ are given by
       \begin{equation}
          {\cal N}_{\pm}^{-2}=2\,\left(1\pm\exp\left(-2\,|\alpha|^{2}
          \right)\left(\frac{1-|\zeta|^2}{1+|\zeta|^2}\right)^{N}\right)\ .
          \label{eq014}
       \end{equation}
The expression (\ref{eq013}) is useful because it allows us to calculate the
expectation values of the relevant operators in closed form, in
particular the expectation value of the Hamiltonian. Evaluating the
energy surface with respect to the symmetry adapted states, one
obtains
	\begin{eqnarray}
	\langle H \rangle_\pm &=& \pm \frac{1}{2} \left(p^2+q^2\right)
	\left\{1-\frac{2}{1 \pm e^{\pm(p^2+q^2)} (\cos\theta)^{\mp
            N}}\right\}\nonumber \\
	&-&\frac{N}{2} \, \omega_{A}
	 \left\{(\cos \theta)^{\pm 1} 
	 \pm \frac{\tan^2\theta \, \cos\theta }{1 \pm e^{\pm(p^2+q^2)} 
	(\cos\theta)^{\mp N} }\right\} \nonumber \\
	&+& \sqrt{2 \, N} \, \gamma  \left\{\frac{\pm p \, \tan\theta \, 
	\sin\phi + q \, e^{p^2+q^2}
	\sin\theta \, \cos \phi \, (\cos\theta)^{-N} }{
	e^{p^2+q^2} (\cos\theta)^{-N}  \pm 1 } \right\} \, .
	\label{symad}
	\end{eqnarray}
In the limit $N \rightarrow \infty$, $\langle H \rangle_+$ reduces to Eq.(\ref{eq003}),
when $\vert\cos\theta\vert\neq 1$. However, it is important to stress
that the analysis can be carried out for any value of $N$, and in this
contribution we work at finite $N$.

\begin{figure}[h]
\scalebox{0.7}{\includegraphics{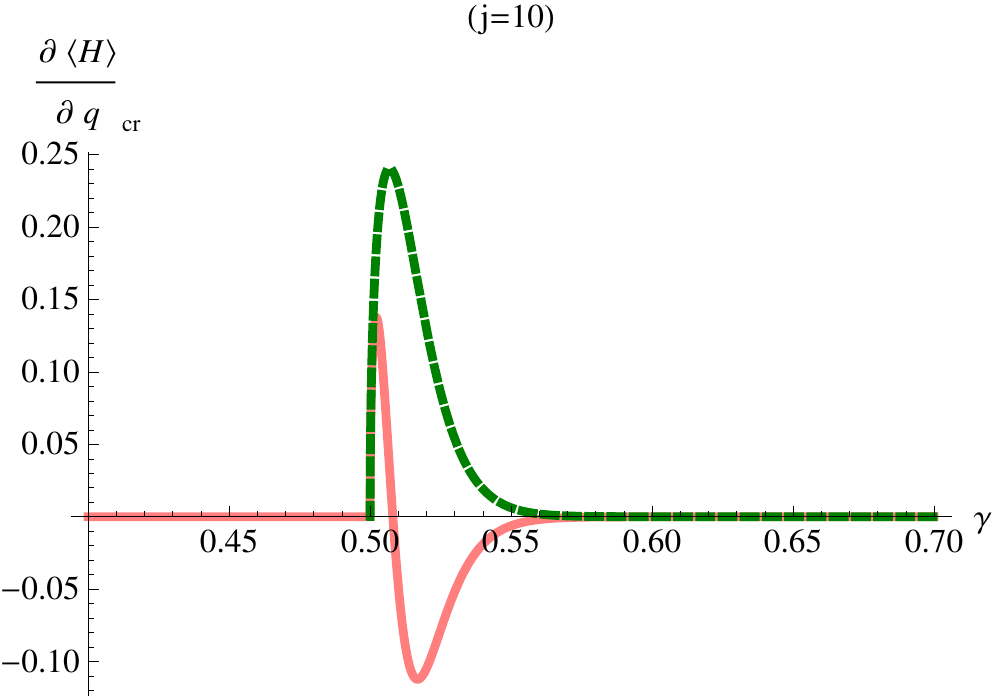}}\qquad
\scalebox{0.7}{\includegraphics{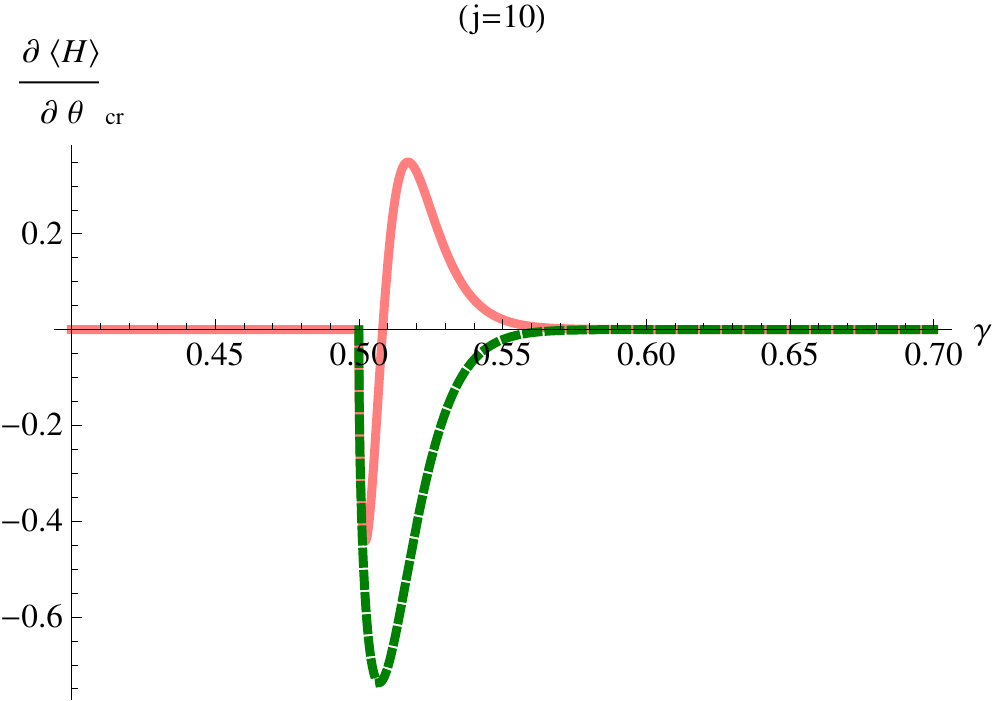}}
\caption{\label{dE_sr}
Derivatives of the energy surface of the symmetry-adapted states, evaluated at the critical points $q_c$ and $\theta_c$, are plotted as a function of the interaction strength $\gamma$. We use 
the frequency $\omega_{A}=1$ and $N=20$ atoms.  Continuous lines refer
to the ground state while dashed lines to the first excited
state.}
\end{figure}

\begin{figure}[h]
\scalebox{0.8}{\includegraphics{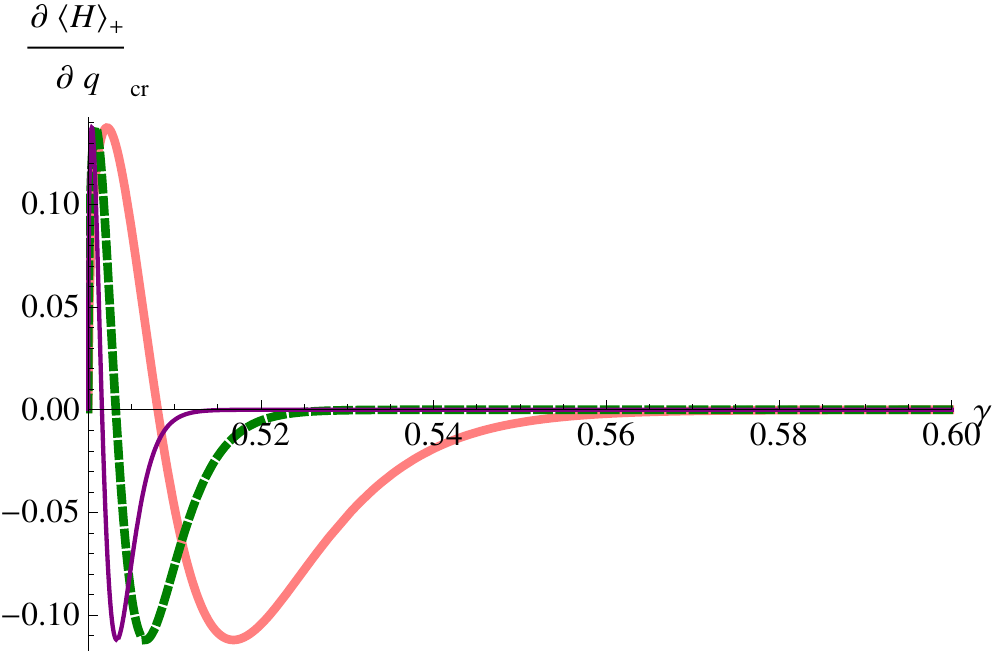}}
\caption{\label{dEq_sr_js}
Derivative of the energy surface of the adapted-symmetry states, calculated at $q_c$ and $\theta_c$, is displayed as a function of the
interaction strength $\gamma$, for different number of atoms. The
plots correspond to $N=20,\, 50,\, 100$ atoms. It can be seen that
the neighborhood of discrepancy from zero diminishes as $N$
increases. In all cases $\omega_{A}=1$.}
\end{figure}

The traditional approach in many-body physics is to use the critical
points in Eq.(\ref{criticos}) of the original energy surface, in order
to obtain the trial state which approximates the two lowest energy
states, and in which to evaluate the expectation values of the
observables.  Formally, one should calculate the critical points of
(\ref{symad}), instead of that, we use the critical points associated 
to Eq.(\ref{eq003}). However all of their critical points also are 
extreme values for (\ref{symad}), except in a small vicinity of
the separatrix, as shown in Fig.(\ref{dE_sr}). In that figure, the derivatives
of the energy surface, given in (\ref{symad}), with respect to $q$ and $\theta$, evaluated at the critical points $q_c$ and $\theta_c$, are
plotted against $\gamma$. (The derivatives with respect to the other
two variables, $p$ and $\phi$, are identically zero). The
neighborhood around the separatrix, where they differ, diminishes as
$j$ increases (cf. Fig.(\ref{dEq_sr_js})). For the superradiant region 
in which the points are critical , they are also minima. The stability analysis of the energy surface (\ref{symad}) and the study of their behaviour in the
near-neighborhood of the separatrix will be reported elsewhere~\cite{castanos2011_03}.

By substituting the critical points
$p_{c}=0$,  $q_{c}=-2\,\sqrt{j}\,\gamma_{c}\,x\,\sqrt{1-x^{-4}}\,\cos\phi_{c}$,
and  $\cos\theta_{c}=x^{-2}$ in (\ref{symad}), the
energy surface associated to the superradiant regime takes the form
   \begin{equation}
	\langle H \rangle_{\pm}=
      -N\gamma_c^2 x^2\,\left[ 2 - (1-x^{-4})\,\frac{1\mp F}{1\pm F}\right]\ ,
      \label{symad2}
   \end{equation}
where we have defined
   \begin{equation}
      x=\gamma/\gamma_{c}\ ,\qquad
      F=x^{-2N}\,\be^{-2N\,\gamma_c^{2}\,x^2\left(1-x^{-4}\right)}\ .
   \end{equation}
The overlap between the ordinary coherent states and our symmetry-adapted states can be written as
	\begin{equation}
		\vert \langle\alpha_c\,\zeta_c\,\vert\,\alpha_c\,\zeta_c\rangle_{\pm}\,\vert^2 = \frac{1}{2}\,(1 \pm F)\ .
		\label{overlapcohsym}
	\end{equation}

\begin{figure}[h]
\scalebox{0.8}{\includegraphics{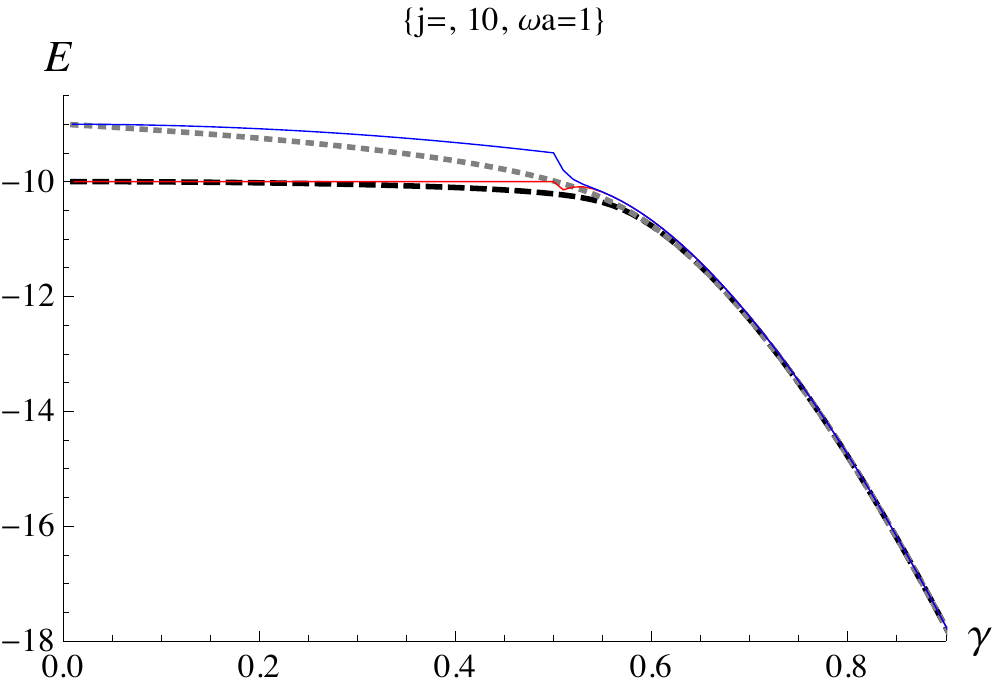}}\qquad\qquad
\caption{\label{energias}
Energies of the ground and first excited states of the Dicke
Hamiltonian as a function of the interaction strength $\gamma$, for
the frequency $\omega_{A}=1$ and $N=20$ atoms.  The thin and thick
continuous lines show the ground and first excited energies,
respectively, evaluated with the projected variational states, while
the dashed and dotted lines correspond to the ground and first excited
energies calculated through the diagonalization of the matrix
Hamiltonian.} 
\end{figure}

In Fig.~(\ref{energias}) the energy surfaces for the even and odd
parities of $\lambda$ are shown together with those for the exact
solution, showing a remarkable agreement between them in spite of
having considered a small number of atoms $N=20$. For the odd parity
case, we have proposed, in the normal region
($\vert\gamma\vert<\gamma_{c}$), a combination of states with
$\lambda=1$:
	\begin{equation}
	   \vert\phi_{1}\rangle=\cos\Omega\,\vert 0\rangle\otimes
               \vert
               j,\,-j+1\rangle-\frac{\gamma}{|\gamma|}\,\sin\Omega\,
               \vert 1\rangle\otimes\vert j,\,-j\rangle\ ,
	   \label{lambda1}
	\end{equation}
and minimized the expectation value of the Hamiltonian with respect to
$\Omega$.  One finds
$\tan(2 \, \Omega_c)=2\,|\gamma|/(1-\omega_{A})$. This gives, for the
resonant case, $\Omega_c=\pi/4$.

\section{Statistical Properties}

\subsection{Expectation values}

One can calculate the expectation values of the matter and field main
operators in the superradiant regime with respect to the symmetry
adapted states. The results are collected in Table~\ref{tab1} together
with those associated to the coherent states. Both are evaluated at
the critical points for the energy surface, and are given in terms of
$x$ and $F$.

	\begin{table}[h!]
	\caption{\label{tab1}
	Expectation values and fluctuations of matter and field
	observables for the coherent and symmetry-adapted states in
	the superradiant regime. The mean-field behaviour obtained in the
	normal region can be recovered by taking the limit
	$x\rightarrow 1$.}
	\begin{center}
	\begin{tabular}{|c|c|c|}
		\hline 
		\raisebox{-1.5ex}[0pt]
		&Coherent
		&Symmetry Adapted\\
		\hline
		\rule[-3mm]{0mm}{5mm}
		$\langle\hat{q}\rangle$
		&$-\sqrt{2N}\,\gamma_{c}\,x\,\sqrt{1-x^{-4}}$
		&$0$\\
		\hline
		\rule[-3mm]{0mm}{5mm}
		$\langle\hat{p}\rangle$
		&$0$
		&$0$\\
		\hline
		\rule[-3mm]{0mm}{5mm}
		$\langle \hat{J}_{x}\rangle$
		&$\frac{N}{2}\,\sqrt{1-x^{-4}}$
		&$0$\\
		\hline
		\rule[-3mm]{0mm}{5mm}
		$\langle \hat{J}_{y}\rangle$
		&$0$
		&$0$\\
		\hline
		\rule[-3mm]{0mm}{5mm}
		$\langle \hat{J}_{z}\rangle$
		&$-\frac{N}{2}\,x^{-2}$
		&$-\frac{N}{2}\,x^{2}\left(1-\frac{1-x^{-4}}{1\pm F}
		\right)$\\
		\hline
		\rule[-3mm]{0mm}{5mm}
		$\langle\hat{a}^{\dagger}\hat{a}\rangle$
		&$N\,\gamma_{c}^{2}\,x^{2}\,\left(1-x^{-4}\right)$
		&$N\,\gamma_{c}^{2}\,x^{2}\,\left(1-x^{-4}\right)\left(\frac{
		1\mp F}{1\pm F}\right)$\\
		\hline
		\rule[-3mm]{0mm}{5mm}
		$\langle\hat{\Lambda}\rangle$
		&$\frac{N}{2}\left(1-x^{-2}+2\,\gamma_{c}^{2}\,x^{2}\,
                        \left(1-x^{-4}\right)\right)$
		&$\frac{N}{2}\left(\frac{1-x^{-2}}{1\pm F}\right)
		\Big\{x^{2}+2\,\gamma_{c}^{2}\,x^{2}\,\left(1+x^{2}\right)$
		\phantom{pittsburg}\\
		&&$\phantom{pittsburgh}\mp\left(
		x^{4}+2\,\gamma_{c}^{2}\,x^{2}\,\left(1+x^{2}\right)\right)
                        \,F\Big\}$\\
		\hline
		\rule[-3mm]{0mm}{5mm}
		$\mathbf{(\Delta\hat{q})^{2}}$
		&$\frac{1}{2}$
		&$\frac{1}{2}+2N\,\gamma_{c}^{2}\,x^{2}\,\left(\frac{
                        1-x^{-4}}{1\pm F}\right)$\\
		\hline
		\rule[-3mm]{0mm}{5mm}
		$(\Delta\hat{p})^{2}$
		&$\frac{1}{2}$
		&$\frac{1}{2}\mp 2N\,\gamma_{c}^{2}\,x^{2}\,\left(
                        \frac{1-x^{-4}}{1\pm F}\right)\,F$\\
		\hline
		\rule[-3mm]{0mm}{5mm}
		$\mathbf{(\Delta\hat{J}_{x})^{2}}$
		&$\frac{N}{4}\,x^{-4}$
		&$\frac{N}{4}\left(1+\frac{\left(N-1\right)\left(
		1-x^{-4}\right)}{1\pm F}\right)$\\
		\hline
		\rule[-3mm]{0mm}{5mm}
		$(\Delta\hat{J}_{y})^{2}$
		&$\frac{N}{4}$
		&$\frac{N}{4}\left(1\pm\frac{\left(N-1\right)
		\left(1-x^{4}\right)F}{1\pm F}\right)$\\
		\hline
		\rule[-3mm]{0mm}{5mm}
		$(\Delta\hat{J}_{z})^{2}$
		&$\frac{N}{4}\,\left(1-x^{-4}\right)$
		&$\frac{N}{4}\,\frac{\left(1-x^{-4}\right)}{\left(1\pm
		F\right)^{2}}\,\left[1\mp\left(N-1\right)\left(1
		-x^{4}\right)\,F-x^{4}F^{2}\right]$
		\\
		\hline
		\rule[-3mm]{0mm}{5mm}
		$(\Delta\,\hat{a}^{\dagger}\hat{a})^{2}$
		&$N\,\gamma_{c}^{2}\,x^{2}\,\left(1-x^{-4}\right)$
		&$N\,\gamma_{c}^{2}\,x^{2}\,\bigg\{N\,\gamma_{c}^{2}
                        \,x^{-6}\,\left(1-x^{4}\right)
		\left(\frac{1\mp F}{1\pm F}\right)^{2}
                        $\phantom{luisoctaviosfmal}\\
		&&$+\left(1-x^{-4}\right)\left[N\,\gamma_{c}^{2}\,x^{2}\,
                        \left(1-x^{-4}\right)
		+\frac{1\mp F}{1\pm F}\right]\bigg\}$\\
		\hline
		\rule[-3mm]{0mm}{5mm}
		$\langle \hat{J}_{z}\,\hat{a}^{\dagger}\hat{a}\rangle$
		&$-N\,\gamma_{c}^{2}\,\left(1-x^{-4}\right)$
		&$-N\,\gamma_{c}^{2}\,x^{4}\left(1-x^{-4}\right)
                        \left(\frac{x^{-4}\mp F}{1\pm F}\right)$\\
		\hline
		\rule[-3mm]{0mm}{5mm}
		$\langle \hat{J}_{x}\,\hat{q}\rangle\rangle$
		&$-\sqrt{\frac{N^{3}}{2}}\,\gamma_{c}\,x\,
		\left(1-x^{-4}\right)$
		&$-\sqrt{\frac{N^{3}}{2}}\,\gamma_{c}\,x\,
                        \frac{1-x^{-4}}{1\pm F}$\\
		\hline
	\end{tabular}
	\end{center}
	\end{table}

\noindent
Observe that, for expectation values different from zero in the symmetry adapted states, the coherent state results can be obtained
from the former by letting $F$ go to zero,
with the exception of $(\Delta\hat{q})^{2}$ and
$(\Delta\hat{J}_{x})^{2}$. These two exceptions arise because
$\langle\hat{q}\rangle=\langle\hat{J}_{x}\rangle=0$ for the former but
not for the latter. That $F$ tends very quickly to zero as a function
of $\gamma$ (or $x$) can be seen in Fig.~(4), especially for large
$N$. This is why coherent states have been so successful in the past
as trial functions.
It can be seen that the quantities which differ the most are some
fluctuations of matter and field observables. To assess this
difference we evaluated the fluctuation of the atomic transition
operator $\hat{J}_{x}$ and of the first quadrature of the electromagnetic field, $\hat{q}$.


\begin{figure}[h]
\scalebox{0.8}{\includegraphics{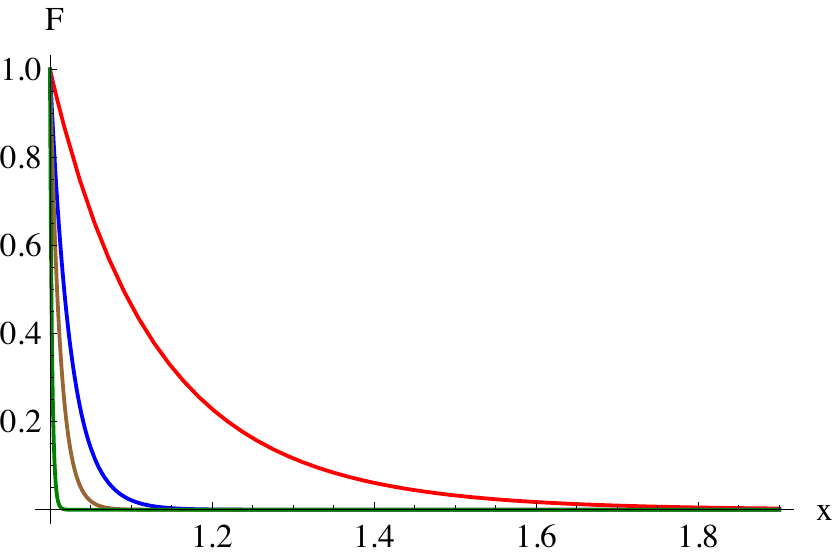}}
\qquad\qquad
\caption{\label{funF}
The behavior of $F$ as a function of $x$ is presented for a number of
atoms in the cavity equal to $N=2,10,20$ and $100$. We use
$\gamma_{c}=0.5$ (in resonance).} 
\end{figure}

\subsection{Fluctuations $\Delta\hat{J}_x$ and $\Delta\hat{q}$}

\begin{figure}[h]
\scalebox{0.5}{\includegraphics{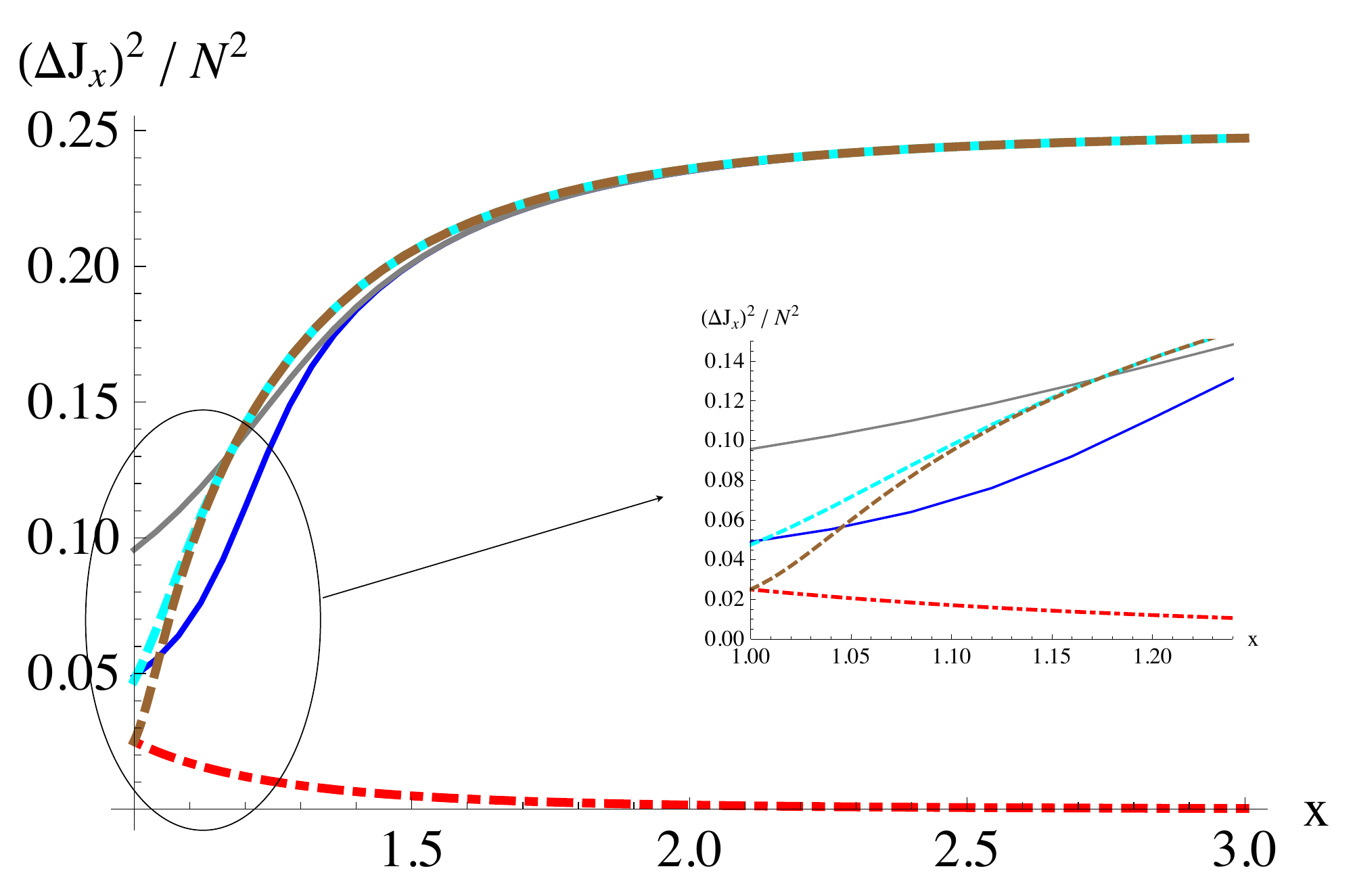}}
\qquad\qquad
\caption{\label{d2jx}
Fluctuation of the atomic transition operator $\hat{J}_{x}$ evaluated
for the ground and first excited states of the Dicke Hamiltonian as a
function of the interaction strength $\gamma$, for the frequency
$\omega_{A}=1$ and $N=10$ atoms.  The dashed dark and light lines show
the fluctuations for the ground and first excited states, respectively, evaluated with
the projected variational states, while the continuous lines describe the correspodning
fluctuations calculated through the
diagonalization of the matrix Hamiltonian. The result for the coherent
state is shown in the dashed-dotted line.(Color online.)}
\end{figure}

\begin{figure}[h]
\scalebox{0.4}{\includegraphics{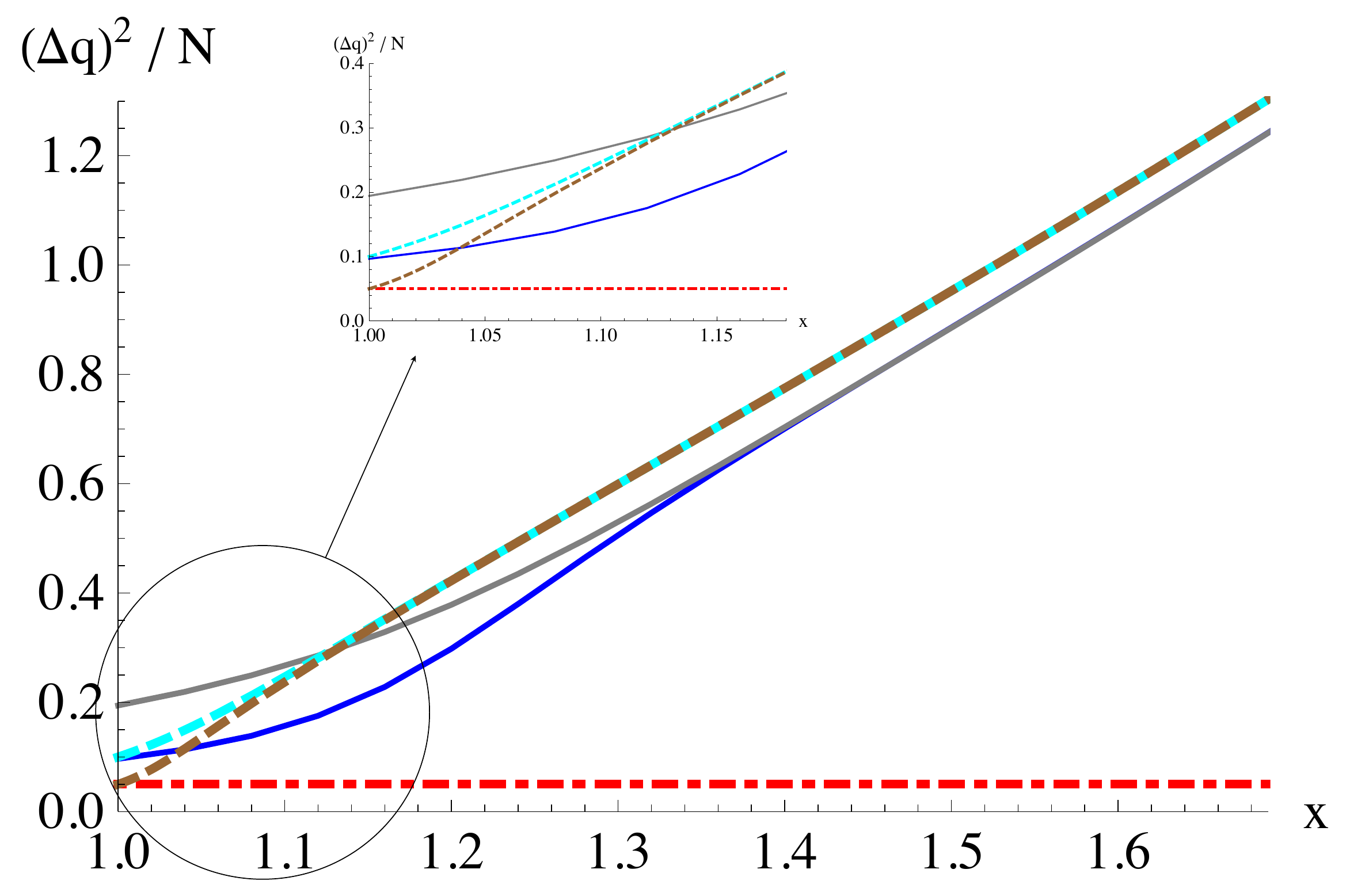}}
\qquad\qquad
\caption{\label{d2q}
Fluctuation of the electromagnetic field first quadrature operator
$\hat{q}$ evaluated for the ground and first excited states of the
Dicke Hamiltonian as a function of the interaction strength $\gamma$,
for the frequency $\omega_{A}=1$ and $N=10$ atoms.  The dashed dark
and light lines show the fluctuations of the ground and first excited states,
respectively, evaluated with the projected variational states, while
the continuous lines display the fluctuations of the ground and first excited
states calculated through the diagonalization of the matrix
Hamiltonian. The result for the coherent state is shown in the
dashed-dotted line.(Color online.)}
\end{figure}

In Fig.~(\ref{d2jx}) we show the fluctuation of the transition
operator $(\Delta \hat{J}_{x})^{2}$ for the ground and first excited
states of the Dicke model with $N=10$ atoms and $\omega_{A}=1$,
calculated for the symmetry adapted states and the coherent
states.
While $(\Delta \hat{J}_{x})^{2}/N^{2}$ tends erroneously to zero for
the coherent state, as $x$ increases, the results for the
symmetry-adapted states show the appropriate quadratic dependence in
$N$, asymptotically reaching $(\Delta \hat{J}_{x})^{2}\approx N^{2}/4$.
In Fig.~(\ref{d2q}) we show $(\Delta \hat{q})^{2}$ for the
ground and first excited states of the Dicke model with $N=10$ atoms
and $\omega_{A}=1$ calculated for the symmetry-adapted states and the
coherent states. While for the coherent state the result is a constant
$1/2$, for the symmetry adapted states one has a linear dependence in
the number of atoms. Additionally, in both cases one can see that the
symmetry-adapted states compare really well with the exact result
obtained from the diagonalization of  Hamiltonian for the even and odd
parity states, except in the close vicinity of the separatrix of the
physical system~\cite{rapcom}.
These results show clearly the benefit of using the symmetry-adapted
states over the coherent ones.

\subsection{Joint probability distribution function}

\begin{figure}[h]
\scalebox{0.7}{\includegraphics{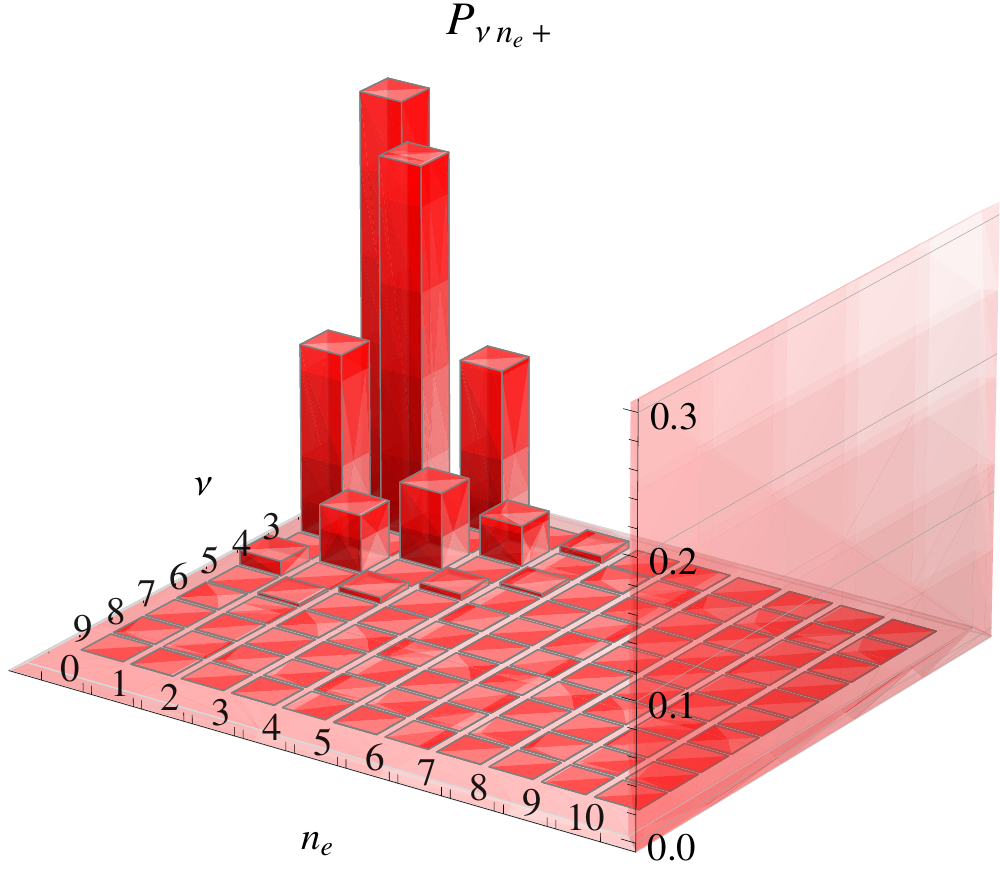}} \qquad
\scalebox{0.7}{\includegraphics{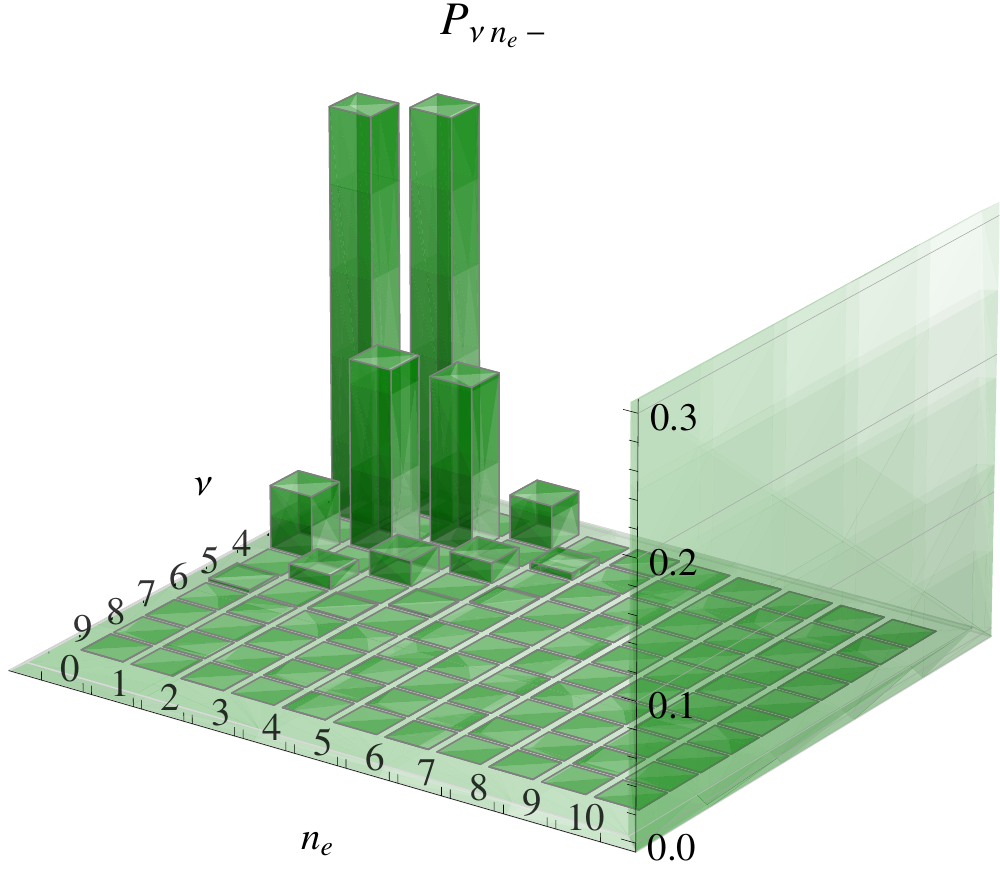}} 
\caption{\label{conjunta055}
The joint probability distribution function of photons and excited atoms is shown, for   
$\gamma_c=0.5$, $\gamma=0.55$, and $N=10$ atoms.  At the left,  
we plot the result for the even parity states  
while at the right we indicate the distribution for the odd ones. }
\end{figure}

\begin{figure}[h]
\scalebox{0.7}{\includegraphics{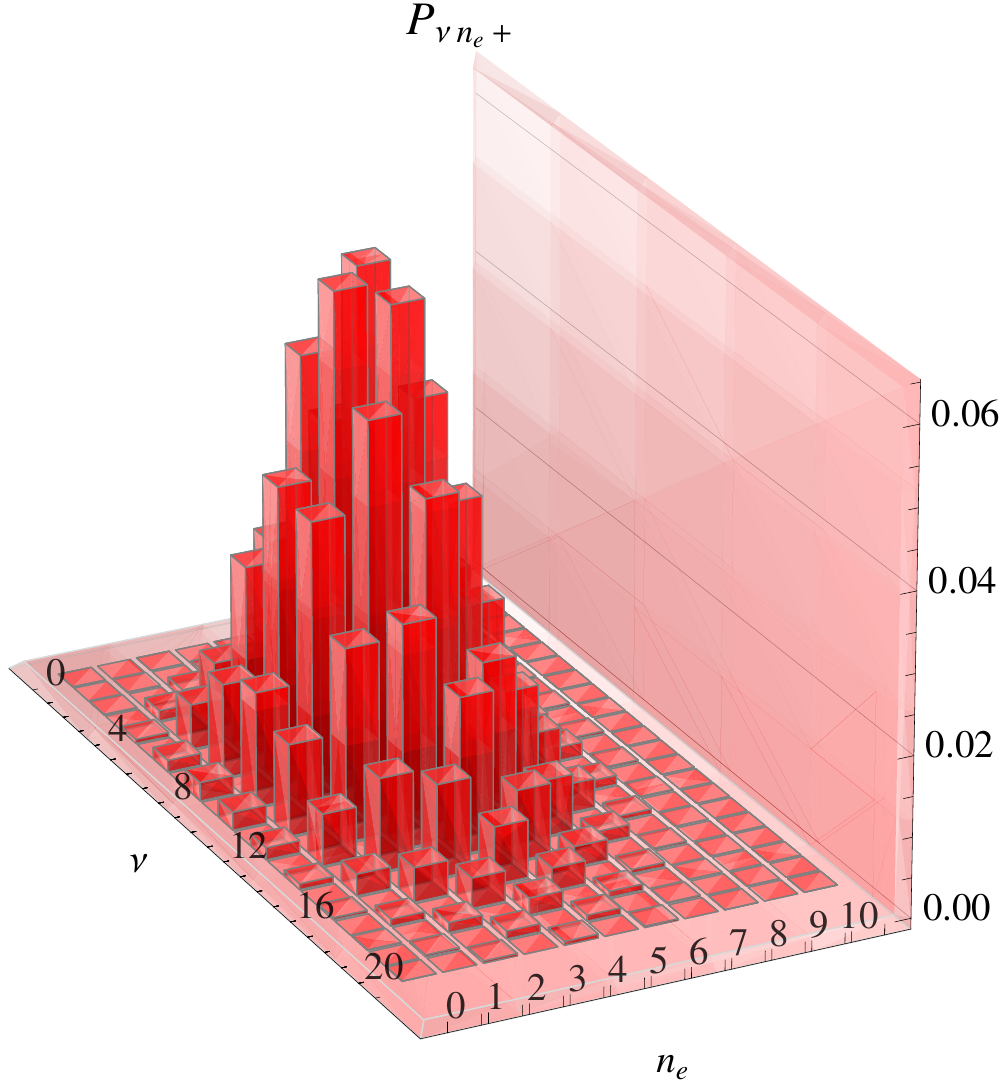}} \qquad
\scalebox{0.7}{\includegraphics{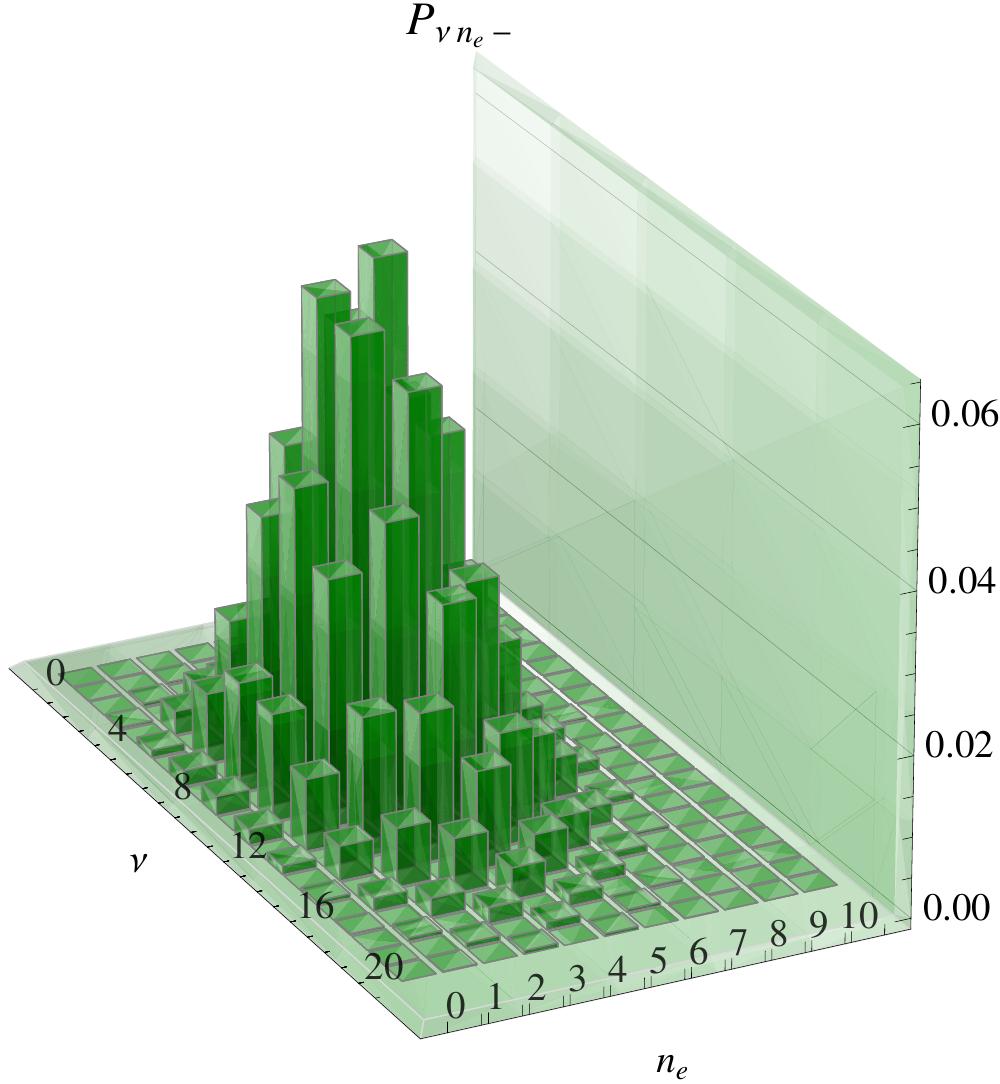}} 
\caption{\label{conjunta100}
The joint probability distribution function of photons and excited
atoms is shown, for
$\gamma_c=0.5$, $\gamma=1.0$, and $N=10$ atoms.  At the left,
we plot the result for the even parity states
while at the right we indicate the distribution for the odd ones.}
\end{figure}

The joint probability of finding $\nu$ photons and $n_e=j +m$ excited
atoms, for even and odd parity states, is obtained by taking the modulus 
square of the scalar product of the Fock and angular momentum states 
$\vert \nu\rangle\otimes\vert
N,\,n_e \rangle$ with  the symmetry-adapted states. The
result depends on the values of $N$, $\gamma_c$, and $x$, and it is
given by
   \begin{eqnarray}
      {\cal P}_{\pm}(\nu,\,n_{e})&=&\left[1\pm\left(-1
      \right)^{\nu+n_{e}}\right]\,\frac{1}{\nu!}
      \left[N\,\gamma_c^{2} \, x^2 \, \left(1-x^{-4}\right)
      \right]^{\nu}\nonumber\\
      & &\times \binom{N}{n_{e}}\left(\frac{1-x^{-2}}{2}
      \right)^{n_{e}}\left(\frac{1+x^{-2}}{2}
      \right)^{N-n_{e}}\, x^N \, \frac{\sqrt{F}}{1\pm F}\ .
   \end{eqnarray}
The joint probability distribution functions, for the even- and
odd-parity states, are shown in Figs.~(\ref{conjunta055},
\ref{conjunta100}) for a system of $N=10$ atoms and in resonance,
i.e., $\omega_{A}=1$. In Fig.~(\ref{conjunta055}), we consider
$\gamma=0.55$ and in Fig.~(\ref{conjunta100}), $\gamma=1.0$. It is
clear that there are holes in the distributions of each plot,
corresponding to the forbidden values (odd or even) of the sum $\nu+n_{e}$.
Note that the Poissonian distribution when near the separatrix
($\gamma=0.55$) turns into a quasi-normal  distribution as we move
away ($\gamma=1$). While very few photons and excited atoms contribute
to the state in the first case, many more contribute in the second
case, as is to be expected.

By summing over the number of photons $\nu$ or the excited atoms
$n_e$, one determines the corresponding marginal distributions. For
the distribution of the number of photons one has
   \begin{equation}
      {\cal P}_{\pm}(\nu)=\frac{1}{\nu!}\left[N\,
      \gamma_c^{2} \, x^2 \,\left(1-x^{-4}\right)\right]^{\nu}\,
      \left[x^{N}\pm\left(-1\right)^{\nu}x^{-N}\right]
      \,\frac{\sqrt{F}}{1\pm F}\ ,
      \label{phdist}
	\end{equation}
while for the distribution of the number of excited atoms the result is
	\begin{equation}
      {\cal P}_{\pm}(n_{e})=\binom{N}{n_{e}}\,\left(
      \frac{1-x^{-2}}{2}\right)^{n_{e}}\,\left(
      \frac{1+x^{-2}}{2}\right)^{N-n_{e}}
      \,\frac{1\pm\left(-1\right)^{n_{e}}x^{2N}F}{
      1\pm F}\ .
   \end{equation}
   
\begin{figure}[h]
\scalebox{0.8}{\includegraphics{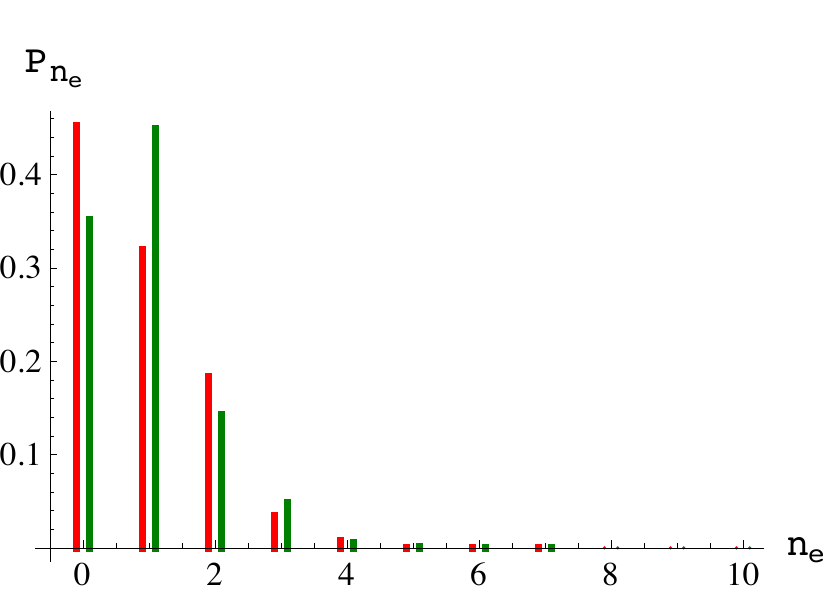}} \quad
\scalebox{0.8}{\includegraphics{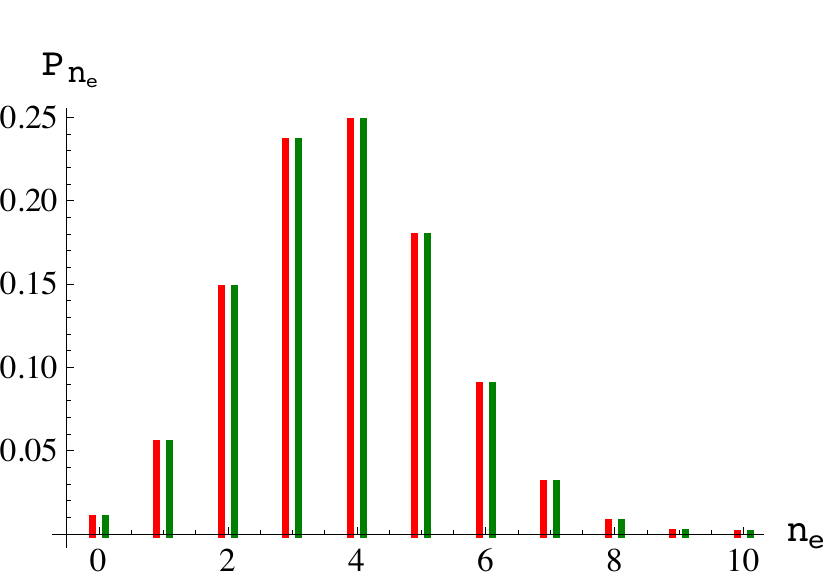}} 
\caption{\label{excited}
The probability distribution of excited atoms is shown, for   
$\gamma_{c}=0.5$ and $N=10$ atoms.  At left we plot the result for
$\gamma=0.55$, while at the right for $\gamma=1$. In both cases the
red bar (left) indicates the distribution of the even parity states and
the green bar (right) the result for the odd ones.}
\end{figure}

\begin{figure}[h]
\scalebox{0.8}{\includegraphics{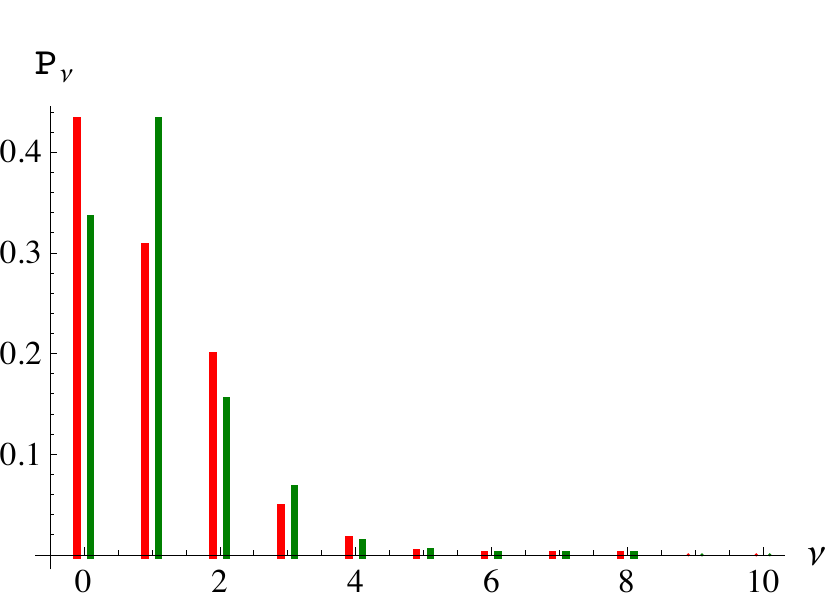}} \quad
\scalebox{0.8}{\includegraphics{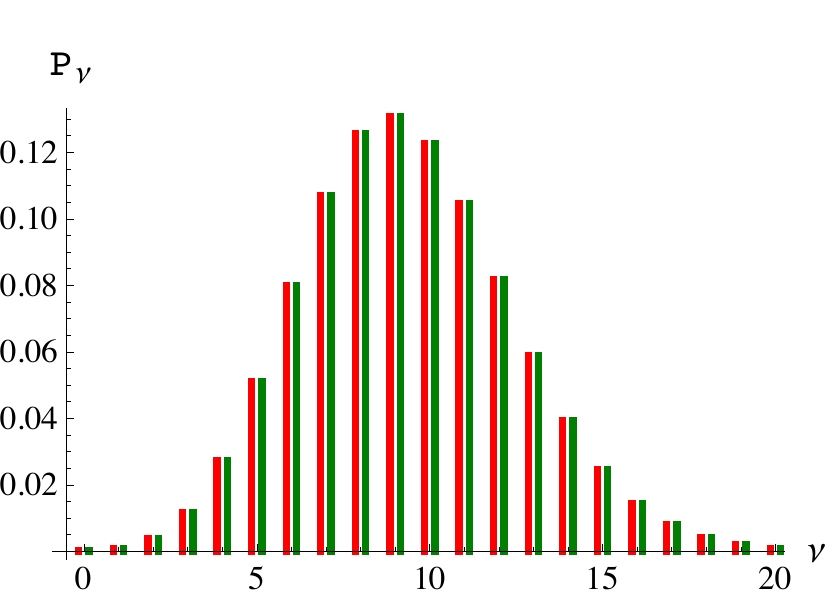}} 
\caption{\label{fotones}
The probability distribution of photons is shown, for   
$\gamma_{c}=0.5$ and $N=10$ atoms. At left we plot the result for
$\gamma=0.55$, while at the right for $\gamma=1$. In both cases the
red bar (left) indicates the distribution of the even parity states and
the green bar (right) the result for the odd ones.}
\end{figure}

The behavior of these probability distribution functions for even and odd parity states is shown in Figs.~(\ref{excited}) and ~(\ref{fotones}) for values of $\gamma$ close to ($x=1.1$) and far from ($x=2.0$ ) the separatrix. 
For $x=1.1$, they are different, while for $x=2$, they cannot be distinguished.
We notice that the binomial and Poissonian distributions become 
quasi-normal as we move away from the separatrix. This is
to be expected, since,
from the results given in Table~\ref{tab1}, it is immediate that the
distribution function of photons, Eq.~(\ref{phdist}), can be rewritten
as
   \begin{eqnarray}
      {\cal P}_{\pm}(\nu)&=&\frac{1}{\nu!}\mu^{\nu}\be^{-\mu}
      \left[\frac{1\pm(-1)^{\nu}x^{-2N}}{1\pm\be^{-2\mu}x^{-2N}}
      \right]\ ,
   \end{eqnarray}
where $\mu=\langle\hat{a}^{\dagger}\hat{a}\rangle$ is the expectation
value of the photon number operator in the coherent state. For large
values of $N$ this distribution becomes a Poisson distribution, which
in this limit is equivalent to the normal distribution
   \begin{eqnarray}
      {\cal P}_{\pm}(\nu)&\approx&\frac{1}{\sqrt{2\pi\mu}}
      \exp\left[-\frac{\left(\nu-\mu\right)^{2}}{2\mu}
      \right]\ .
   \end{eqnarray}

In Fig.~(\ref{excited}), the distribution of the number of excited
atoms for the even- and odd-parity states are shown. For $x=1.1$,
close to the separatrix, they are different, while for $x=2$, we can not
distinguish them.

For the distribution function of the excited number of atoms,
in the limit when $N\gg 1$ one has
   \begin{eqnarray}
      {\cal P}_{\pm}(n_{e})&\approx&\frac{1}{\sqrt{\pi
      \left(1-x^{-4}\right)N/2}}
      \exp\left[-\frac{\left(n_{e}-\left(1-x^{-2}\right)N
      /2\right)^{2}}{\left(1-x^{-4}\right)N/2}
      \right]\ ,
   \end{eqnarray}
which for large values of $x$ also gives a normal distribution:
   \begin{eqnarray}
      {\cal P}_{\pm}(n_{e})&\approx&\frac{1}{\sqrt{\pi N/2}}
      \exp\left[-\frac{\left(n_{e}-N/2\right)^{2}}{N/2}
      \right]\ .
   \end{eqnarray}

\section{Projected States vs. Exact Solution}

By substituting the critical points of the superradiant phase into the
symmetry adapted variational states~(\ref{eq013}, \ref{eq014}), we
obtain
   \begin{eqnarray}
      \vert\alpha_{sr}, \zeta_{sr}\rangle_{\pm}&=&
      \frac{\pm 1}{\sqrt{2^{N+1}\left(1\pm F\right)}}\,
      \exp\left[-\tfrac{N}{2}\gamma_{c}^{2}\,x^{2}\,\left(1-x^{-4}\right)
      \right]\,\sum_{\nu,\,n_{e}}\left(1\pm(-1)^{\nu+n_{e}}\right)\,
      \frac{\left(\sqrt{N}\,\gamma_{c}\,x\right)^{\nu}}{\sqrt{\nu!}}
      \,\nonumber\\
      &\times& \binom{N}{n_{e}}^{1/2}\,\left(1-x^{-2}\right)^{(\nu+n_{e})/2}\,
      \left(1+x^{-2}\right)^{(N+\nu-n_{e})/2}\,
      \vert\nu\rangle\otimes\vert \tfrac{N}{2},\,n_{e}-\tfrac{N}{2}\rangle\ .
   \end{eqnarray}
In the limit $x\gg 1$ we have $F\rightarrow 0$,
$\alpha_{sr}\rightarrow\sqrt{N}\,\gamma_{c}\,x$, and
$\zeta_{sr}\rightarrow 1$. This implies that each matter coherent
state goes to an eigenfunction of $J_{x}$, i.e.,  a rotation by
${\pi\over 2}$ acting on the state $\vert j,\,-j \rangle$. In that limit the Dicke model becomes completely integrable~\cite{rapcom}.

A good measure of the distance between quantum mechanical states is
given by the fidelity; for pure quantum states it measures their
distinguishability in the sense of statistical
distance~\cite{wooters}, but it is customary to use the fidelity as a
{\it transition probability} regardless of whether the states are pure
or not.

\begin{figure}[h]
\scalebox{0.35}{\includegraphics{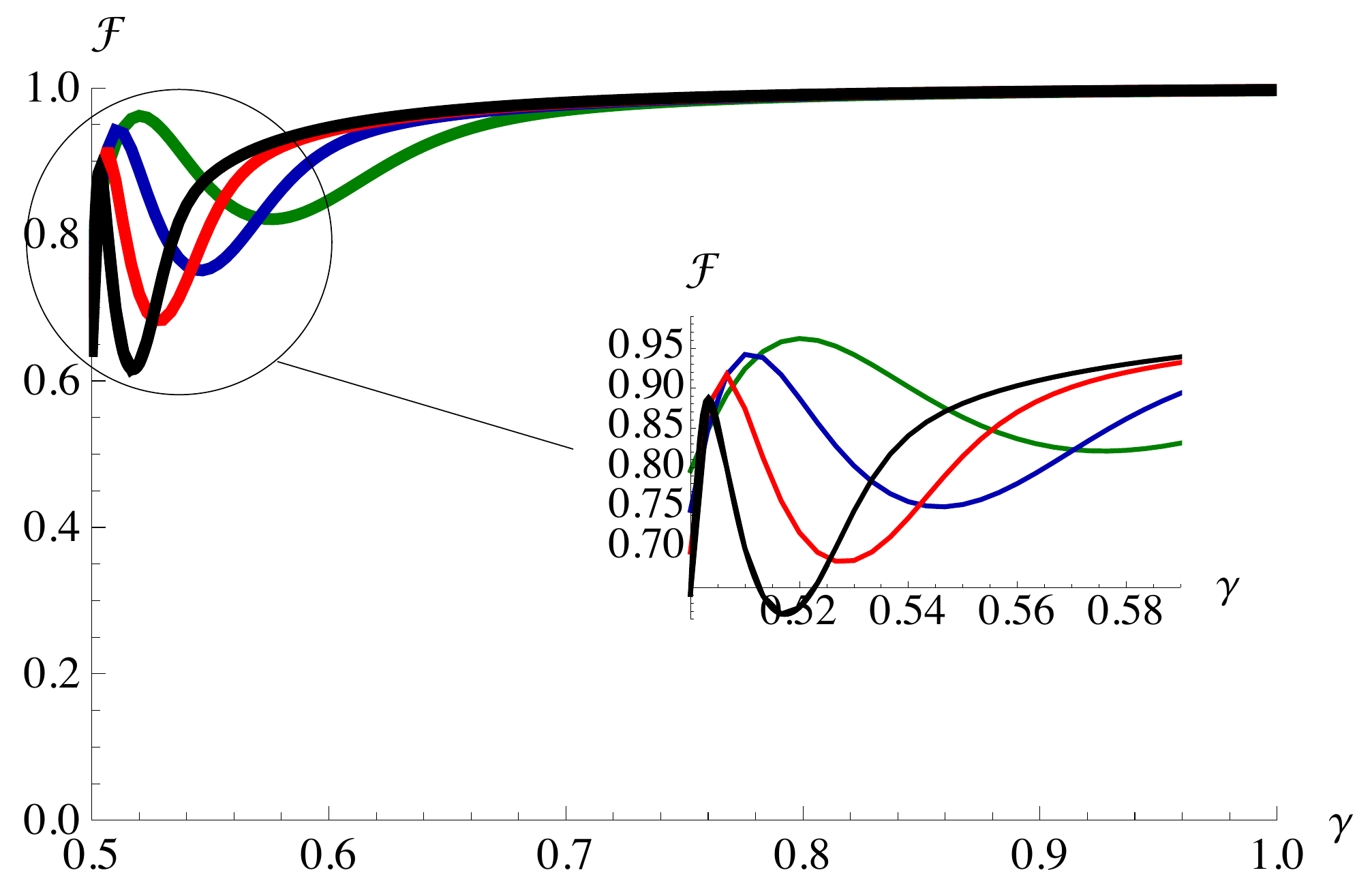}} \qquad
\scalebox{0.35}{\includegraphics{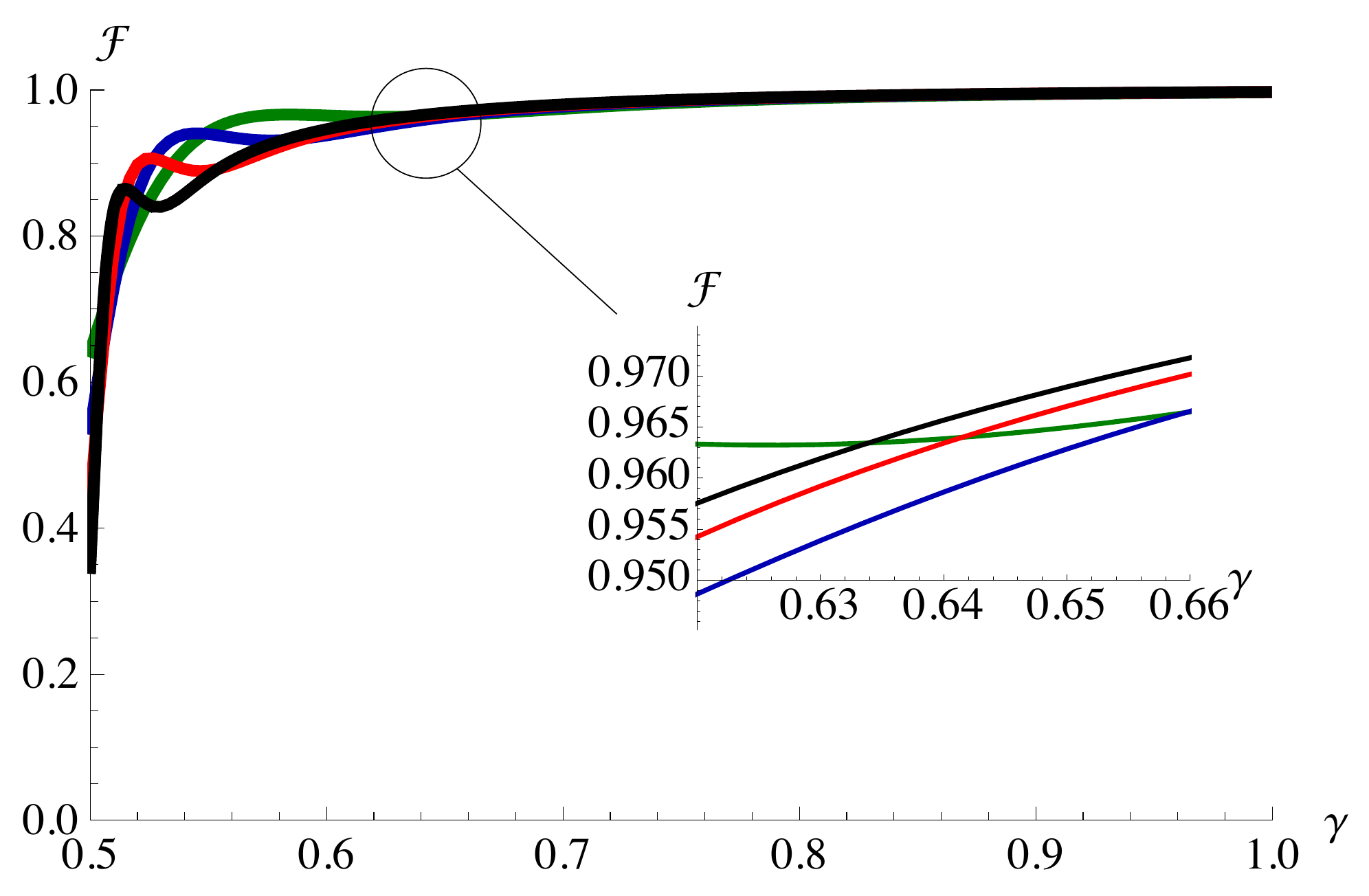}}
\caption{\label{fidelity}
Fidelity between the symmetry-adapted and exact quantum states, as a
function of the interaction strength $\gamma$, shown for the ground
state (even parity, left) and the first excited state (odd parity,
right). In both cases, the plots show that for increasing
$N=10,\,20,\,40$ and $50$ the fidelity rises more sharply to $1$ as we
move away from the phase transition (at $\gamma_c=0.5$). In all cases
$\omega_{A}=1$. (Color online.)}
\end{figure}

In this contribution we calculate the fidelity of the exact ground
and first excited states with respect to the corresponding symmetry
adapted states
   \begin{equation}
       {\mathcal F} = \left|\langle\psi_{P}\vert\psi_{\hbox{exact}}\rangle
       \right|^{2}\ . 
       \label{fidelityec}
   \end{equation}
In both cases the fidelity gives a result very close to 1, except in
the vicinity of the quantum phase transition.  In
Fig.(\ref{fidelity}), the fidelity is shown as a function of $\gamma$
for $N=10,\,20,\,40$ and $50$. For increasing $N$ the fidelity rises
more sharply to $1$ as we move away from the phase transition, located
at $\gamma_{c}=0.5$. Note that (cf. Eq.(\ref{overlapcohsym})), had we used the coherent states as trial functions, we would at best obtain a fidelity $\mathcal{F}$ close to $1/2$, in the limit of large $\gamma.$

\section{Discussion and Conclusions}

By means of the symmetry properties of the Dicke Hamiltonian, we found
analytic expressions for the ground and first excited states, which 
allows also to determine in closed form the expectation values of 
matter and field observables.
The procedure was the following: we calculated the expectation value of the
Hamiltonian with respect to the tensorial product of Weyl and SU(2)
coherent states. This expectation value defines a function called
{\em energy surface} depending on phase space variables and
parameters. We then determine the degenerate and minimum critical
points of the energy surface. The minimum critical points yield the
expressions~(\ref{ecoherente}) which give the mimimum energy of the
system together with information about the quantum phase transition
present in the Dicke Hamiltonian. This quantum phase transition
occurs when
$\tilde{\omega}_{A}\,\tilde{\omega}_{F}=4\tilde{\gamma}^{2}$. Afterwards we restore the symmetry exhibited for the Dicke Hamiltonian by means of the projection of the states to definite parity of the excitation number operator. These symmetry considerations lead to determine the ground and first excited states together with important differences from the mean field results for some fluctuations of the matter and field observables. Additionally they provide us with the joint probability distribution functions of the number of photons and excited atoms.

The
condition to transit from the normal to the superradiant regimes is very difficult to satisfy for optical systems because the
available dipole coupling strengths are usually smaller than the
transition frequency. Some proposals to overcome these problems have
been discussed in~\cite{nagy}, where it is stated that the quantum motion
of a Bose-Einstein condensate trapped in an optical cavity can be used
to realize the Dicke model. Indeed, this has been reported
in~\cite{baumann}, where the Dicke Hamiltonian has been physically
realized in a superfluid gas moving in an optical cavity. However, the feasibility of reaching the superradiant phase of the field-matter interaction is still under debate~\cite{nataf,viehman}.

The
predicted $\sqrt{N}$--behaviour for the collective $N$--atom
interaction strength of the model has been observed experimentally in circuit QED for up to $5$ qubits~\cite{fink}.

In this work we have shown that semi-classical states adapted to the 
symmetry of the Hamiltonian are an
excellent approximation to the exact quantum solution of the ground
and first excited states of the Dicke model in the superradiant phase. Their overlap to the exact quantum states is very close to $1$ except in a close vicinity of the quantum phase transition (cf. Fig.(\ref{fidelity})), whereas that of the ordinary coherent states would be at best equal to $1/2$ (cf. Eq.(\ref{overlapcohsym})). Our projected states
have analytical forms in terms of the model parameters and allow us to
calculate analytically the expectation values of field and matter
observables. We have found that in the superradiant regime the fluctuation 
$(\Delta \hat{q})^2$ of the first quadrature of the electromagnetic field is different 
from the results obtained via the standard coherent states and its value grows 
as a linear function of $N$. Something similar happens for the fluctuation in the dipole transition operator $(\Delta \hat{J}_x)^2$, where one finds a quadratic dependence in the number of atoms. Both these results contradict those obtained previously~\cite{emary}. The expectation values of the number of photons and of the number of excited atoms were also studied, finding that there are no singularities at the phase transition. Those found previously~\cite{nagy} are an artifact of an inappropriate truncation of the Hamiltonian.   The joint probability distribution functions for the ground and first excited states were shown, which may be used to characterize the states of atoms in a cavity according to whether $n_e + \nu$ has an even or odd value.

In a future contribution, we will evaluate other
properties like the entanglement entropy between field and matter, and
the squeezing parameter for the electromagnetic field and atomic
components, which has been the subject of much interest~cite{vidal}.

Finally we want to remark that the present formalism allows for a simplification of 
the exact quantum calculation by considering an expansion of Dicke and photon
number states running from a minimum value $\lambda_{min}$ to
$\lambda_{max}$, where these values can be estimated from
$\lambda_{c}$ and its corresponding fluctuations $\delta\lambda_{c}$
as functions of $\gamma$, the coupling parameter between the field and
matter. The same simplification can immediately be done in the
proposed state~(\ref{eq013}) without changing the obtained results for
all the expectation values and probability distributions.


\section{Acknowledgments}
This work was partially supported by CONACyT-M\'exico
(project-101541), FONCICYT (project-94142), and DGAPA-UNAM.


\end{document}